\documentstyle[epsfig]{aipproc}
\title{Removing Instrumental Artifacts: Suspension Violin Modes}
\author{Soma Mukherjee$^{(1),(2)}$ and Lee Samuel
Finn$^{(1),(2),(3)}$}
\address{$(1)$ Center for Gravitational Physics and Geometry, The
Pennsylvania State University; $(2)$ Department of Physics, The
Pennsylvania State University; $(3)$ Department of Astronomy and
Astrophysics, The Pennsylvania State University.}
\begin{document}
\maketitle
\begin{abstract}
  We describe the design of a Kalman filter that identifies suspension
  violin modes in an interferometric gravitational wave detectors data
  channel.  We demonstrate the filter's effectiveness by applying it
  to data taken on the LIGO~40M prototype.
\end{abstract}

\section{Motivation}

The wire suspensions of the interferometer test masses are a conduit
for environmental noise to enter the gravitational wave data channel;
additionally, they are a source of noise themselves.  Both the
pendulum mode (whose frequency is out of the interferometers band) and
the suspension wire violin modes (whose fundamental mode frequencies
are in-band) are energized by their contact with the
thermal bath.  Owing to their weak damping this energy is
strongly concentrated about the mode resonant frequencies.  At the
fundamental violin mode frequencies the thermal noise dominates the
other noise sources by approximately 50~dB. In addition to this
thermal noise component, non-thermal excitations of the suspension
wires (e.g., sudden creep events) can lead to excitations in the
interferometer output.

These narrow band features are instrumental artifacts, not
gravitational waves: along with other artifacts they should be removed
from the data before it is studied for the presence of other signals. 
As instrumental artifacts, however, they carry important information
about the instrument's state.  Thermal and technical noise that
disturbs the suspension excites these modes and move the mirrors,
leading to an artifact in the gravitational wave channel. 
Gravitational waves give rise to a signal by changing the distance
between the mirrors, but do not move or otherwise excite the
suspension modes.  Correspondingly, if we can determine the mode state
as a function of time we have a way of eliminating a broad class of
technical noise sources that might otherwise masquerade as
gravitational wave bursts.

How can we identify the state of the suspension violin modes, given
only the gravitational wave channel? This is a classic problem in data
analysis, generally addressed by a Kalman filter. We have developed
such a filter for use in LIGO and shown, using data taken in November
1994 at the LIGO 40M prototype, that it enables us to follow the state
of the violin modes independently of the other noise sources --- both
technical and fundamental --- that contribute to the gravitational
wave channel.

\section{The Kalman Filter}

The Kalman Filter \cite{kalman60a} is a mechanism for
predicting the multi-dimensional {\em state\/} of a dynamical system from
a multi-dimensional {\em
  observable.} The system is assumed to evolve linearly and the
observable is assumed to be linearly related to the state.  denoting the
system state
$\mathbf x$ we have (for discrete time
series):
\begin{equation}
{\mathbf x}[k] = {\mathbf A}\cdot{\mathbf x}[k-1] + {\mathbf w}[k-1]. 
\end{equation}
We assume that the system is driven by a stochastic force, referred to as
the {\em process noise\/} and denoted $\mathbf w$. The state dynamics
determine the linear operator
${\mathbf A}$. 

The state contributes to the observation $\mathbf y$ , which also includes
a
  stochastic, additive {\em measurement noise\/} $\mathbf v$:
\begin{equation} 
{\mathbf y}[k]={\mathbf C}\cdot{\mathbf x}[k] + {\mathbf v}[k]. 
\label{eq:meas} 
\end{equation} 
In the classic Kalman filter the process and measurement noises are
assumed to be Normal processes with known co-variances $\mathbf W$ and
$\mathbf V$.\footnote{Even when $\mathbf w$ and $\mathbf v$ are
  not Normal the Kalman filter estimates of the state ${\mathbf x}[k]$ can
be shown to have
  the smallest mean-square error of all linear state estimators that
  depend only on the co-variances.}

Now suppose that we have an estimate $\widehat{\mathbf x}[k-1]$ of the
state, and also an estimate of the error co-variance ${\mathbf
  P}[k-1]$ in the estimate, at sample $k-1$.  The Kalman filter uses
these estimates, the observation ${\mathbf y}[k]$ at sample $k$, and
$\mathbf A$, $\mathbf C$, $\mathbf W$ and $\mathbf V$ to form an
estimate of the state and its error co-variance at sample $k$:
\begin{eqnarray}
  \widehat{\mathbf x}[k] &:=& {\mathbf K}[k]\cdot\left(
    {\mathbf y}[k] - \widehat{\mathbf y}[k]
  \right)\\
  \widehat{{\mathbf P}}[k] &:=&
  \left({\mathbf I}-{\mathbf K}[k]\cdot{\mathbf C}\right)
  \cdot\widetilde{{\mathbf P}}[k]\cdot
  \left({\mathbf I}-{\mathbf K}[k]\cdot{\mathbf C}\right)^{T}
\end{eqnarray}
where
\begin{eqnarray}
  \widehat{\mathbf y}[k] &:=& {\mathbf C}\cdot{\mathbf
    A}\cdot\widehat{\mathbf x}[k-1]\\ 
  {\mathbf K}[k] &:=& 
  \widetilde{{\mathbf P}}[k]\cdot{\mathbf C}^T / 
  \left({\mathbf V} + 
    {\mathbf C}\cdot\widetilde{{\mathbf P}}[k]{\mathbf C}^T
  \right)\\
  \widetilde{\mathbf P}[k] &:=& {\mathbf
    A}\cdot\widehat{\mathbf P}[k-1]\cdot{\mathbf A}^T + {\mathbf W}.
\end{eqnarray}
The estimated system state ${\mathbf x}[k]$ ({\em e.g.,} the generalized
coordinate and conjugate momentum of the violin mode normal mode of the
wire) is thus completely determined by the observation
${\mathbf y}[k]$, the estimated state at sample
$k-1$, the wire dynamics, and the statistical properties of the
process and measurement noise. The error in the estimate ${\mathbf x}[k]$
falls with $k$, converging upon a
limiting error covariance that is fully determined by ${\mathbf A}$,
${\mathbf
  C}$, ${\mathbf W}$ and ${\mathbf V}$; correspondingly, we can choose
any initial estimate of ${\mathbf x}$ and $\mathbf P$ and the filter
will, after several iterations, adjust the state estimate and error
accordingly.

From the state estimate at each sample we can, through the measurement
equation, estimate the contribution of the system to the actual
observation. This estimated contribution can be subtractively removed
from the actual observation, leaving a residual that is as free from
the contaminating influence of the process as we can make it.

\section{Modeling the violin modes}
To describe a Kalman filter for our system we need a model for the
wire state dynamics, the relationship between the wire state and the
appearance of the mode in the detector data channel, and estimates of
the process and measurement noise.

The wire motion contributes significantly to the data channel
only in a narrow band about the violin mode resonant frequency;
correspondingly, we can focus attention this narrow band and model the
wire dynamics as a viscously damped harmonic oscillator driven by white 
noise\footnote{Over the relevant, narrow band near the resonant
peak there is no distinction between viscous and structural damping.}:
\begin{equation}
\ddot\psi = - \omega_0^2 \psi - {\gamma}\dot\psi +
N(t).\label{eq:dyn}
\end{equation} 
We assume that the measurement is of the state variable $\psi$ plus
white measurement noise. 

There are many violin modes, corresponding to the many wires that are
used to suspend the interferometer mirrors: for each wire there are
separate state variables and a separate equation describing the
dynamics of that mode.

Using standard lock-in techniques we mix the $\psi$ and the data channel
with a local oscillator whose frequency is near that of the violin modes
and band-limit the output of the lock-in to the narrow band over which
our model is accurate. The in-phase and quadrature-phase components of
mixed-down $\psi$ become the state ${\mathbf x}$ used in the Kalman
filter, and the in-phase and quadrature-phase components of the data
channel become the observation ${\mathbf y}$. The matrices ${\mathbf
  A}$ and $\mathbf C$, describing the state dynamics and the
relationship between the state $\mathbf x$ and observation $\mathbf y$
are derived from the our model in light of the lock-in,
band-limiting, and discrete time sampling operations.

\section{Results and Discussion}

\subsection{Removing the artifact}

To explore the effectiveness of the Kalman filter in identifying the
contribution of the violin modes to the detector output we have
applied it to data taken in November 1994 at the LIGO~40M prototype
detector. In this instrument the fundamental violin mode resonances
are all in the  (571.6, 605.425)~Hz band. 

\begin{figure}[h]
\epsfysize=0.5\textwidth
\begin{center}
\leavevmode\epsffile{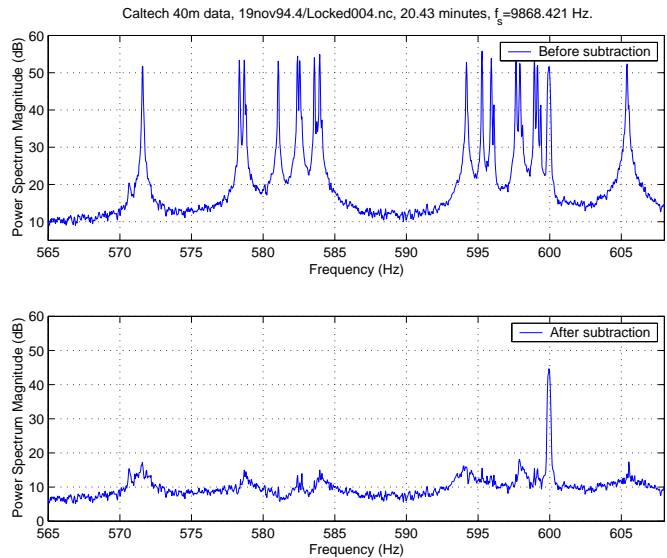}
\end{center}
%\centerline{\epsfig{file=JUL8_powersub.eps,width=5.75in,height=3.9in}}
  \caption{Power spectrum of the 40 meter data
  showing the violin modes between 571 and 605 Hz before and after
  subtraction of the Kalman estimates for all the modes.  The
  remaining line feature is the 9th harmonic of the 60~Hz power main. 
  Features like these are dealt with in other ways
  \protect\cite{sintes98a,finn99h,finn99i,allen99c}.}\label{fig:40power} 
\end{figure}

The upper and lower panels of figure \ref{fig:40power} show the power
spectra of the interferometer output in a 45~Hz band between 565.0 and
610.0~Hz before and after the subtractive removal of the Kalman
filter estimate of the violin mode contribution to the detector
output. The filter identifies the contribution of the mode to the
detector output, allowing us to suppress this artifact by 40~dB.  The 
residual bumps positioned in the wings of the removed lines, are 
non-linear artifacts: the violin mode amplitudes are so 
large that they modulate the detector transfer function, up-converting
other detector noise frequency-modulating the violin mode signal itself.

\subsection{Statistics of the artifact and residual}

Some simple exploratory statistics show the value of identifying and
removing the known instrumental artifacts from the data stream.

\begin{figure}[h]
\epsfxsize=0.5\textwidth
\begin{center}
\epsffile{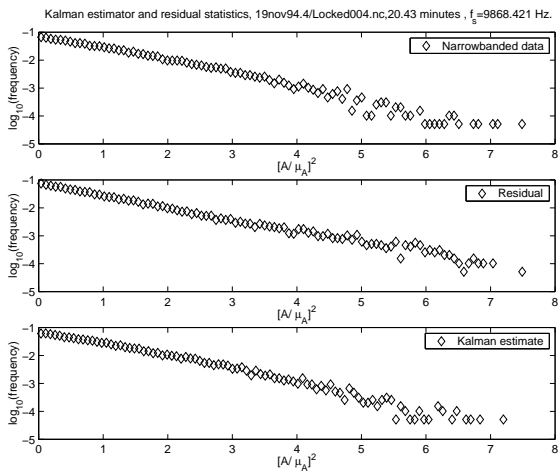}
\end{center}
%\centerline{\epsfig{file=JUL8_statall.eps,width=5.75in,height=3.5in}}
\caption{Histograms of the square of Kalman
  estimated amplitude divided by mean amplitude for (top) data in the
  narrow-band (570-599.5 Hz) having the fundamental violin modes,
  (middle) residual after subtraction of the Kalman estimates of the
  modes from the data and (bottom) the Kalman estimated violin 
modes. Deviations from Gaussianity occur with much
lower frequency in the residual than in the mode
contribution to the observation.}\label{fig:40expstat}  
\end{figure}

Figure~\ref{fig:40expstat} shows a histogram of the sample amplitude
relative to the RMS sample amplitude for the data channel
before (top) and after (middle) removal of the Kalman filter estimate
of the violin mode contribution, and (bottom) for the estimated violin
mode contribution itself. In an ideal world the measurement noise and
the process noise are Normal; correspondingly, each of these
distributions should be Rayleigh and appear, in these figures, as
straight lines.  Departures from a straight line thus imply
non-Gaussian noise statistics. 

Comparing the three panels in figure~\ref{fig:40expstat} shows 
that the violin mode artifact contributes significantly to the
non-Gaussian component of the noise in the detector data
channel. Since gravitational waves do not excite the violin modes,
this excess noise component is strictly technical. Generalizing to the
full-scale detector, removing artifacts like these, together with
their associated excess noise, from the data stream before analysis
thus strengthens our ability to make significant statements regarding
the detection of gravitational waves in the residual.

\subsection{Violin modes vs. gravitational waves}

The Kalman filter estimates the violin mode contribution to the
gravitational wave channel.  If that estimate is influenced by the
presence of a gravitational wave signal then removing the estimated
contribution may distort evidence of the wave in the output.

The Kalman filter identifies the violin mode contribution through its
dynamics.  Since the evolution with time of expected gravitational
wave signals is different than the dynamics of the mode contribution
to the detector output, we expect that the Kalman estimates of the
violin mode contribution will not be influenced by the presence of a
gravitational wave signal.

It is useful to consider two different kinds of sources: burst
sources, such as inspiraling neutron star binary systems, and periodic
sources, such as a pulsar at a frequency near to but not identical
with the violin mode frequency.

Nearly all the 
signal to noise ratio  signal-to-noise from an inspiraling binary
is contributed when the signal is in the band from about 70~Hz to 250~Hz
\cite{finn96a,finn97g}: what happens in the band near the violin mode
is is inconsequential.  Similarly, the filter output is entirely
unaffected by what happens in the band where the S/N is deposited. 
We have verified this by forming estimates of the violin
mode state from the LIGO~40M prototype data, and from the same data set
but with an added, simulated gravitational wave signal
corresponding to a coalescing neutron star binary.  Even for a very
strong signal there is no difference in the predicted mode
amplitudes when the  signal was in the relevant band.

In separate experiments we have looked at how well the Kalman filter 
rejects nearby monochromatic signals, such as might arise from a pulsar. 
As long as the signal is greater than a linewidth away from 
the violin mode, the estimated state is unaffected by the periodic
signal;  correspondingly, the periodic signal is unaffected 
the subtractive removal of the estimated contribution of the mode to the
output. 

\subsection{Monitoring the wire state} 

The Kalman filter estimates separately the state of each violin mode;
correspondingly, we can monitor the state of each of these wires
separately. It is convenient to represent the state in terms of its
amplitude and phase (as opposed to generalized coordinate and
momentum); additionally, it is convenient to express the amplitude as
an instantaneous measure of the energy in the mode, expressed in terms
of temperature. Excess noise will raise the effective temperature of the
mode, while the projection of the corresonding mirror motion normal to
the optical axis will lower the effective temperature. 

\section{Summary}

Thermal and technical noise that disturbs the violin modes of the
substrate suspension moves the mirrors, leading to a strong,
narrow band artifact in the gravitational wave channel.  Gravitational
waves, on the other hand, also change the distance between the mirrors,
but without moving or otherwise exciting these modes.  The
Kalman filter described here distinguishes between excitations due to
gravitational waves and those due to thermal or other
excitations of the violin modes, allowing us to eliminate a broad class
of technical noise sources that might otherwise masquerade as
gravitational wave bursts.

A Kalman filter uses the known dynamics of the modes to
distinguish between the mode ``signal'' and other
contributions to the measured detector output: {\em i.e.,} it {\em
detects\/} the violin modes.  This distinguishes it from other methods
({\em e.g.,} multitaper methods, linear notch filters
\cite{allen98a}) which purport to characterize or remove artifacts, but
which in fact simply suppress all contributions to the noise within a
narrow band, not distinguishing violin mode from other contributions.

The computational cost of identifying and removing the violin modes
using the Kalman filter described here is negligible: an interpreted
Matlab \cite{matlab} implementation on a low-end  workstation
runs at greater than 20$\times$ the detector's real-time sample
rate.  A compiled implementation, with attention paid to optimization,
an additional  speed-up of 10 or more can be expected. 

We thank Albert Lazzarini for drawing our attention to
Kalman Filtering and the LIGO Laboratory for its hospitality during the
1997/8 academic year and for the use of the 40 meter prototype data. SM
thanks S.~Mohanty for many valuable insights and LSF thanks P.~Fritschel
for valuable discussions. This work was supported by NSF awards PHY
98-00111 and 99-96213.

%\bibliographystyle{prsty}
%\bibliography{phyjabb,keys,references}

\begin{thebibliography}{1}
\bibitem{matlab}
Matlab, a technical computing environment for high-performance numeric
  computations in linear algebra, is a product of {T}he {M}ath{W}orks, Inc.
\end{thebibliography}

\end{document}